\documentclass[runningheads]{llncs}
\usepackage[T1]{fontenc}

\usepackage[utf8]{inputenc}
\usepackage{amsmath,amscd}
\usepackage{amsfonts}
\usepackage{amssymb}
\usepackage{graphicx,color}
\usepackage{bbm}

\begin{document}

\title{Ternary cellular automata induced by semigroups of order 3 are solvable}
\author{Henryk Fuk\'s}

%
%
\author{Henryk Fuk\'s\inst{1}\orcidID{0000-0001-5855-3706}}
\authorrunning{H. Fuk\'s}
%
\institute{Brock University, Department of Mathematics and Statistics, St. Catharines, Canada 
\email{hfuks@brocku.ca}}
\maketitle              

\begin{abstract}
The minimal number of inputs in the local function of a non-trivial cellular automaton  is two.
Such a function can be viewed as  as a kind of binary operation. If this operation
is associative, it forms, together with the set of states, a semigroup. There are 
18 semigroups of order 3 up to equivalence, and they define 18 cellular automata rules
with three states. We investigate these rules with respect to solvability and 
show that all of them are solvable, meaning that the state of a given cell after $n$
iterations can be expressed by an explicit formula. We derive the relevant formulae
for all 18 rules using some additional properties possessed  by particular semigroups of order 3, such as commutativity and idempotence.
\keywords{ternary cellular automata, semigroups, solvability}
\end{abstract}

\section{Introduction}
Cellular automata (CA) are characterized by two crucial parameters, the number of states 
of the alphabet and the  number of inputs of the local function. It is a well known fact
that in the case of one-dimensional CA these two are not completely independent. The number of states of a cellular automaton can be decreased at the expense of increased neighbourhood size \cite{vollmar2013algorithmen,Hadeler2017}, and conversely, the number of inputs can be decreased at the expense of enlarging
the alphabet \cite{Hadeler2017}. The smallest possible non-trivial alphabet has two elements, 
and for this reason a significant part of research on CA is devoted to binary rules.
If one wishes to  study the binary rules in a systematic way, a natural strategy would be to look at rules with increasing neighbourhood size $n$, starting from $n=1, 2, 3 \ldots$.
Binary CA with one or two inputs are not very interesting, yet those with three inputs (called \emph{elementary rules}) already exhibit a rich variety of dynamical behaviour, 
and they have been extensively studied. In particular, it has been recently demonstrated \cite{book1} that
about 65\%  of all elementary rules are \emph{solvable}, meaning that
one can construct an explicit formula for the state of a given cell after a given number of iterations
if the initial configuration is provided. 

Given the aforementioned trade-off between the number of states $n$ and the number of inputs $k$
one could equally well consider CA with the minimal non-trivial neighbourhood size, which
obviously means $n=2$. Systematic study of two-input rules would then involve
investigating those with increasing number of states $k$, starting with $k= 1, 2, 3, \ldots$.
The $k=1$ case is not interesting, and $k=2$ (binary) case is already well known. There are namely 
$2^4=16$ binary two-input rules, and all of them are equivalent to elementary CA with
``effective'' dependence on only two inputs. For example, for the elementary rule 34 the local function $f(x_1,x_2,x_3)=x_3-x_2x_3$ depends only on $x_2$ and $x_3$, thus it is effectively two-input. Since all  elementary rules have been  extensively studied, we will not discuss two-input binary rules here.

The next case, namely 3-state (or \emph{ternary}) two-input rules, are much more interesting.
In general, ternary CA rules have not received much attention in the literature,
although some results regarding their solvability in the probabilistic sense \cite{paper57,paper66} and existence of
rules with conservation laws \cite{WOLNIK2020180} are known.
 There are $3^9=19683$ two-input ternary rules, definitely too large a number to investigate systematically on one-by-one basis. We will, therefore, in this paper look only at a particular class of those rules,
with much smaller number of members. Their solvability will be our main focus.
\section{Definitions}
In order to define a CA rule we need the set of states 
or alphabet $\cal A$ and the local function $f$. In the case of two-input rules the local function is  a function $f: {\cal A}^2 \to \cal A$. A bi-infinite sequence  $x \in {\cal A}^{\mathbbm{Z}}$ of symbols
of the alphabet 
will be called a \emph{configuration}.
With the local function $f$
we associate a global function $F: {\cal A}^{\mathbbm{Z}} \to {\cal A}^{\mathbbm{Z}}$, such that
$$
[F(x)]_i = f(x_i, x_{i+1})
$$
for all $ i \in \mathbbm{Z}$. Multiple iterates of $F$ will be denoted by $F^n$.
Using the traditional nomenclature of cellular automata, $[F^n(x)]_i$ represents the state of the cell  $i$  after $n$ iterations of rule $F$ (with local function $f$).

We will also define a family of functions $f^n :{\cal A}^{n+1}\to \cal A$, such that
\begin{align*}
f^2(x_0,x_1,x_2)&=f\big(f(x_1,x_2), f(x_2,x_3)\big)\\
f^3(x_0, x_1,x_2,x_3)&=f^2\big(f(x_0,x_1), f(x_1,x_2), f(x_2,x_3)\big),\\
& \ldots
\end{align*}
and by induction
$$
f^n(x_0,x_1,\ldots, x_{n})=f^{n-1}\big(f(x_0,x_1), f(x_1,x_2), \ldots,  f(x_{n-1},x_{n})\big).
$$
It is not hard to see that
$$
[F^n(x)]_i = f^n(x_i, x_{i+1}, \ldots, x_{n}).
$$
This means that if we find an explicit formula for $f^n(x_0, x_1, \ldots, x_{n})$
in terms of its $n$ arguments, we will automatically have the formula for the state of cell $i$ after $n$ iterations.

Given a local function with two inputs $f(x_1,x_2)$ one can think of it as a kind of binary operation $\odot$,  $$x_1 \odot x_2=f(x_1,x_2).$$
If $\cal A$ is the alphabet of the CA, then obviously $x_1 \odot x_2 \in \cal A$ for $x_1, x_2 \in \cal A$,
meaning that the operation $\odot$ is closed. If no further conditions are imposed,
the pair $(\cal A, \odot)$ is called \emph{magma}. The set of local functions of ternary CA with two inputs is, therefore, equivalent to the set of magmas with three elements.

If for all $x_1, x_2, x_3 \in \cal A$
$$ (x_1 \odot x_2) \odot x_3 = x_1 \odot (x_2 \odot x_3),$$
then $\odot$ is called \emph{associative}. Associative magmas are called \emph{semigroups}.

A semigroup with $k$ elements is called a semigroup of order $k$. For small~$k$'s, semigroups of order $k$ have been enumerated, most recently up to  $k=10$ \cite{distler2012}. The list of
semigroups of order 3 is know since 1940's, and it is known that there are 113 of them. If we count
semigroups related by isomorphism or anti-isomorphism as  equivalent, then their number reduces to only 18 non-equivalent ones~\cite{distler2010}. Assuming that the semigroup elements are $\{x,y,z\}$, the ``multiplication tables'' of the corresponding operation $\odot$ are shown in Table~\ref{listsg}.
\begin{table}[t]
$$
\begin{array}{ccc}
   G_{1}: \; \ \begin{tabular}{c|ccc}   
  &    x  &  y &   z\\ \hline    
x &    x  &  y &   z\\ 
y &    y  &  z &   x\\ 
z &    z  &  x &   y \end{tabular}   
&
G_{2} : \; \ \begin{tabular}{c|ccc}    
&    x  &  y &   z\\ \hline
x &    y  &  z &   y\\ 
y &    z  &  y &   z\\ 
z &    y  &  z &   y \end{tabular}   
&
G_{3} : \; \ \begin{tabular}{c|ccc}   
  &    x  &  y &   z\\ \hline
x &    y  &  z &   z\\ 
y &    z  &  z &   z\\ 
z &    z  &  z &   z \end{tabular}   
\\[3em]
G_{4} : \; \ \begin{tabular}{c|ccc}   
  &    x  &  y &   z\\ \hline
x &    z  &  y &   x\\ 
y &    y  &  y &   y\\ 
z &    x  &  y &   z \end{tabular}   
&
G_{5} : \; \ \begin{tabular}{c|ccc}   
  &    x  &  y &   z\\ \hline
x &    z  &  x &   x\\ 
y &    x  &  y &   z\\ 
z &    x  &  z &   z \end{tabular}  
&
G_{6} : \; \ \begin{tabular}{c|ccc}   
  &    x  &  y &   z\\ \hline
x &    z  &  x &   x\\ 
y &    x  &  z &   z\\ 
z &    x  &  z &   z \end{tabular}  
\\[3em]
G_{7} : \; \ \begin{tabular}{c|ccc}   
  &    x  &  y &   z\\ \hline
x &    z  &  z &   z\\ 
y &    z  &  z &   z\\ 
z &    z  &  z &   z \end{tabular}  
&
G_{8} : \; \ \begin{tabular}{c|ccc}   
  &    x  &  y &   z\\ \hline
x &    z  &  z &   z\\ 
y &    z  &  y &   z\\ 
z &    z  &  z &   z \end{tabular}  
&
G_{9} : \; \ \begin{tabular}{c|ccc}   
  &    x  &  y &   z\\ \hline
x &    z  &  y &   z\\ 
y &    y  &  y &   y\\ 
z &    z  &  y &   z \end{tabular}  
\\[3em]
G_{10} : \; \ \begin{tabular}{c|ccc}   
  &    x  &  y &   z\\ \hline
x &    z  &  x &   z\\ 
y &    x  &  y &   z\\ 
z &    z  &  z &   z \end{tabular}  
&
G_{11} : \; \ \begin{tabular}{c|ccc}   
  &    x  &  y &   z\\ \hline
x &    z  &  z &   z\\ 
y &    y  &  y &   y\\ 
z &    z  &  z &   z \end{tabular}  
&
G_{12} : \; \ \begin{tabular}{c|ccc}   
  &    x  &  y &   z\\ \hline
x &    z  &  z &   z\\ 
y &    x  &  y &   z\\ 
z &    z  &  z &   z \end{tabular}  
\\[3em]
G_{13} : \; \ \begin{tabular}{c|ccc}   
  &    x  &  y &   z\\ \hline
x &    x  &  y &   z\\ 
y &    y  &  y &   z\\ 
z &    z  &  z &   z \end{tabular}  
&
G_{14} : \; \ \begin{tabular}{c|ccc}   
  &    x  &  y &   z\\ \hline
x &    x  &  z &   z\\ 
y &    z  &  y &   z\\ 
z &    z  &  z &   z \end{tabular}  
&
G_{15} : \; \ \begin{tabular}{c|ccc}   
  &    x  &  y &   z\\ \hline
x &    x  &  x &   x\\ 
y &    y  &  y &   y\\ 
z &    x  &  x &   z \end{tabular}  
\\[3em]
G_{16}: \; \ \begin{tabular}{c|ccc}   
  &    x  &  y &   z\\ \hline
x &    x  &  x &   z\\ 
y &    y  &  y &   z\\ 
z &    z  &  z &   z \end{tabular}  
&
G_{17} : \; \ \begin{tabular}{c|ccc}   
  &    x  &  y &   z\\ \hline
x &    x  &  x &   x\\ 
y &    y  &  y &   y\\ 
z &    z  &  z &   z \end{tabular}  
&
G_{18}: \; \ \begin{tabular}{c|ccc}   
  &    x  &  y &   z\\ \hline
x &    x  &  x &   x\\ 
y &    y  &  y &   y\\ 
z &    x  &  y &   z \end{tabular}  
\end{array}
$$
\caption{List of multiplication tables of 18 semigroups of order 3 (up to equivalence).}\label{listsg}
\end{table}
Each of these multiplication tables define one ternary 2-input CA rule.
We will denote the local function corresponding to $G_i$ by $f_i$.
Let us consider, for example, the semigroup $G_4$. Its multiplication table given in Table~\ref{listsg} with $x=0, y=1, z=2$ becomes
$$
 \begin{tabular}{c|ccc}
 & 0  &  1 &   2\\ \hline 
0& 2  &  1 &   0\\ 
1&  1  &  1 &   1\\ 
2&  0  &  1 &   2
\end{tabular}      
$$
The corresponding local function is given by
\begin{equation}\label{rule4cases}
f_4(u,v)=\begin{cases}
2 & \text{if $(u,v)=(0,0)$ or $(2,2)$,}\\
1 & \text{if $u=1$ or $v=1$,}\\
0 & \text{otherwise.}
\end{cases}
\end{equation}
Note that instead of the representation $x=0, y=1, z=2$, we could have
used other permutation of symbols, for instance, $x=2, y=1, z=0$ etc.
This would result in an equivalent rule but with differently named symbols.

We will now construct the so-called \emph{polynomial representation} for each of the 
18 rules, using the method outlined in \cite{book1}.
This will be a polynomial with two variables $u,v$ 
returning the same values as $f(u,v)=u \odot v$ for all $u, v \in \cal A$.
Define indicator function 
$$
I_n(x)=
\begin{cases}
1 & \text{if $x=n$,}\\
0 & \text{otherwise.}
\end{cases}
$$
The following equation is obviously true
\begin{equation}\label{polyrep}
f(u,v)= \sum _{i,j \in \cal A} f(i,j)I_i(u) I_j(v).
\end{equation}
For the ternary alphabet ${\cal A} = \{0,1,2\}$, the following polynomial forms
of indicator functions can be used if $x \in \cal A$,
\begin{align}\label{indicators}
I_0(x)&= \frac{1}{2}(x-1)(x-2), \nonumber\\
I_1(x)&= x(2-x),\\
I_2(x)&=\frac{1}{2}x(x-1). \nonumber
\end{align}
These have been obtained by direct fitting. For example, for $I_0$ we have three
values needed, $I_0(0)=1$, $I_0(1)=0$ and $I_0(2)=0$, thus we need a polynomial with three 
coefficients, i.e., the quadratic function. Taking $I_0(x)=a x^2+bx +c$ we
have three free constants and three conditions, so we can determine $a$, $b$ and $c$,
resulting in $I_0(x)=\frac{1}{2}x^2-\frac{3}{2}x+1$. This factorizes to 
$I_0(x)=\frac{1}{2}(x-1)(x-2)$, as listed above.
Using expressions given in eq. (\ref{indicators}) to replace 
indicator functions on the right hand side of eq. (\ref{polyrep}), one obtains
a polynomial in two variables representing a given CA. For example, for
$G_4$ the local function $f_4$ given by eq. (\ref{rule4cases}) returns 2 if $(u,v)$ is $(0,0)$ or $(2,2)$,
and it returns 1 if at least one of $(u,v)$ is 1, therefore
\begin{align*}
f_4(u,v)&=2 I_0(u)I_0(v)+2 I_2(u)I_2(v)\\
&+ I_1(u)I_0(v)+ I_1(u)I_1(v) + I_1(u)I_2(v)
+I_0(u)I_1(v)
+I_2(u)I_1(v).
\end{align*}
This simplifies to 
\begin{align*}
f_4(u,v)&=2 I_0(u)I_0(v)+2 I_2(u)I_2(v)
+ I_1(u)+ I_1(v) -I_1(u)I_1(v),
\end{align*}
where we used the fact that $I_0(x)+I_1(x)+I_2(x)=1$. Using the expressions for indicator functions this
yields
\begin{align*}
f_4(u,v)&= \frac{1}{2}(u-1)(u-2)(v-1)(v-2)+\frac{1}{2}u(u-1)v(v-1)\\
&+ u(2-u) + v(2-v) -u(2-u)v(2-v),
\end{align*}
and, after simplification,
$$f_4(u,v)=1+(v-1) (u-1).$$
\begin{table}
\begin{center}
\begin{tabular}{l|c|l}
semigr. & representation & local function \\ \hline
$G_{1}$  & $x=0, y=1, z=2$ & $f_{1} (u,v)= u+v+\frac{3}{4}uv ( 3 uv-5 u-5 v+7)      $      \\
         &                 & $f_{1} (u,v)= (u+v) \mod 3                             $      \\
$G_{2}$  & $x=1, y=2, z=0$ & $f_{2} (u,v)=  1+(v^2-v-1) (u^2-u-1)                   $      \\
$G_{3}$  & $x=1, y=2, z=0$ & $f_{3} (u,v)=   2 u (u-2) v (v-2)                      $      \\
$G_{4}$  & $x=0, y=1, z=2$ & $f_{4} (u,v)=   1+(v-1) (u-1)                          $      \\
$G_{5}$  & $x=0, y=1, z=2$ & $f_{5} (u,v)=  2-(u+v-3)(u v-u-v)                      $    \\
$G_{6}$  & $x=0, y=1, z=2$ & $f_{6} (u,v)=   1+(v^2-3 v+1) (u^2-3 u+1)              $      \\
$G_{7}$  & $x=0, y=1, z=2$ & $f_{7} (u,v)=2                                         $      \\
$G_{8}$  & $x=2, y=1, z=0$ & $f_{8} (u,v)=   u (u-2) v (v-2)                        $      \\
$G_{9}$  & $x=0, y=1, z=2$ & $f_{9} (u,v)=   \frac{1}{4} u v (v-3) (u-3)            $      \\
$G_{10}$ & $x=2, y=1, z=0$ & $f_{10} (u,v)= u v (u+v-u v)                           $      \\
$G_{11}$ & $x=0, y=1, z=2$ & $f_{11} (u,v)= 1+(u-1)^2                               $      \\
$G_{12}$ & $x=2, y=1, z=0$ & $f_{12} (u,v)= u v (2-u)                               $      \\
$G_{13}$ & $x=0, y=1, z=2$ & $f_{13} (u,v)=u+v-\frac{1}{2} u v (u v-2 u-2 v+5)      $      \\
         &                 & $f_{13} (u,v)= \max(u,v)                               $      \\
$G_{14}$ & $x=2, y=1, z=0$ & $f_{14} (u,v)=  \frac{1}{2} u v (3 u v-5 u-5 v+9)      $      \\
$G_{15}$ & $x=0, y=1, z=2$ & $f_{15} (u,v)=   \frac{1}{2}u (u v^2-u v-v^2-2 u+v+4)  $      \\
$G_{16}$ & $x=2, y=1, z=0$ & $f_{16} (u,v)= 1+(v-1)^2 (u-1)                         $      \\
$G_{17}$ & $x=0, y=1, z=2$ & $f_{17} (u,v)= u                                       $      \\
$G_{18}$ & $x=0, y=1, z=2$ & $f_{18} (u,v)=  \frac{1}{2} u (u v-2 u-v+4)            $      \\
\end{tabular}
\end{center}
\caption{Local function of CA defined by semigroups of order 3.}\label{tablelocfunct}
\end{table}

Table \ref{tablelocfunct} shows polynomial representations of all 18 rules
defined by semigroups of order 3, obtained in a similar way as above. As mentioned, when constructing the rule corresponding to a given semigroup, one can choose among $3!=6$ permutations of symbols $\{0,1,2\}$
to represent $\{x,y,z\}$. This will produce six different polynomial representations.
In Table \ref{tablelocfunct} for each semigroup we listed the shortest and simplest polynomial 
among the six, giving in the second column the corresponding choice of $\{x,y,z\}$.
In cases of $G_1$ and $G_{13}$, in addition of the polynomial representation, we also listed  alternative expressions.
\begin{figure}
\begin{center}
\includegraphics[width=12cm]{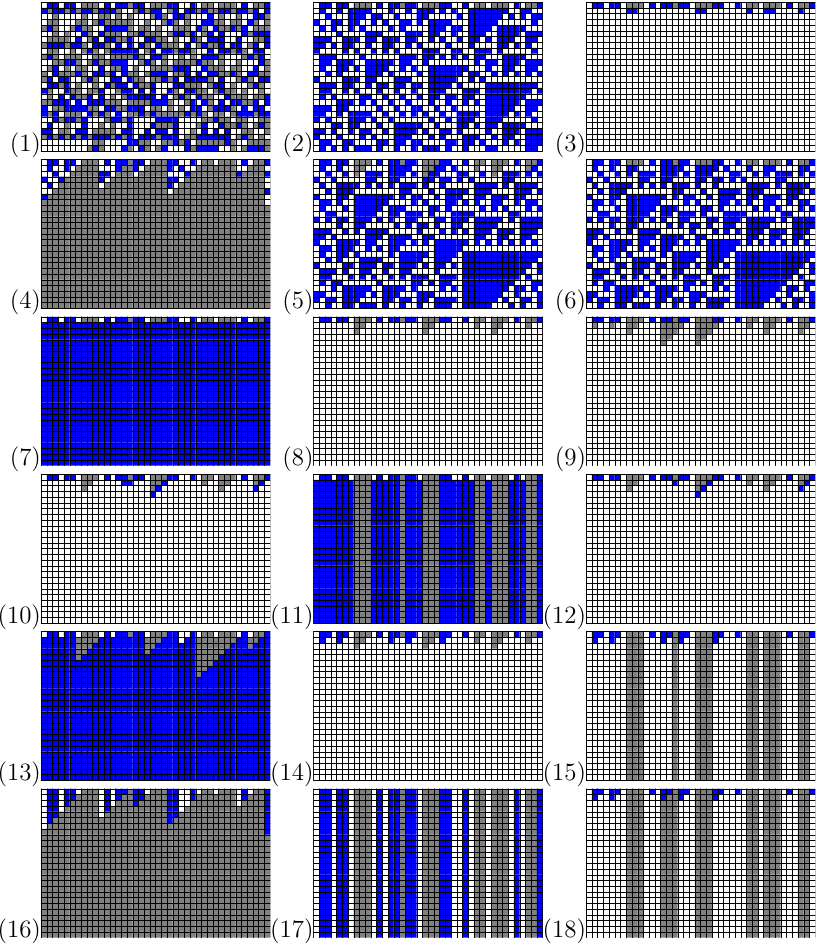}
\end{center}
\caption{Spatiotemporal patterns of all 18 semi-group rules generated from random initial condition of 40 sites  with periodic boundary conditions, iterated 25 times. Color scheme: $0=$white, $1=$gray and $2=$blue.}\label{patterns}
\end{figure}

It is imperative to stress here that the polynomial expressions defining   local functions in the last column of  Table \ref{tablelocfunct}
use regular arithmetic addition and multiplication in the field of rational numbers.
Although the set $\{0,1,2\}$ with addition and multiplication
modulo 3 forms a finite field, we are \emph{not} using this fact in subsequent considerations (except to provide an  alternative definition of $G_1$ and the corresponding solution formula). Polynomials in  Table \ref{tablelocfunct} are simply integer-valued polynomials which take values in $\cal A$
if their arguments belong to~$\cal A$.

In order for the reader to get some idea regarding dynamical behaviour of our 18 CA rules,
Figure~\ref{patterns} show examples of spatiotemporal patterns generated by them,
starting from ``random'' initial conditions with periodic boundaries. We can see that these patterns
range in complexity from very simple to rather complicated ones, the latter exhibiting  
triangular regions of varying sizes with fractal-like scaling often found in 
many other cellular automata. Using informal Wolfram
classification \cite{RevModPhys.55.601}, rules 1, 2, 5 and 6 are class 3, 
rules 11, 15, 17 and 18
are class 2, ad the remaining ones are class 1.

\section{Basic results}
Some semigroups listed in Table \ref{listsg} have additional properties which will be
useful in what follows. Some are \emph{commutative}, meaning that $u \odot v=
v \odot u$ for all $u,v \in \cal A$. Others are \emph{idempotent}, meaning that 
$u \odot u=u$ for all $u \in \cal A$.
Commutative semigroups are $G_1,G_{2}, \ldots, G_{10}$, as well as $G_{13}$ and  $G_{14}$.
Semigroups $G_{13}, G_{14} \ldots,G_{18}$ are idempotent. Semigroups $G_{11}$ and $G_{12}$ 
are neither commutative nor idempotent.

The simplest to deal with are idempotent semigroups. For them it is very straightforward
to construct an explicit formula for $f^n$. 
\begin{proposition}\label{thmidem}
If the semigroup $({\cal A},\odot)$  is idempotent and $f(x_0,x_1)=x_0 \odot x_1$, then 
$$f^n(x_0,x_1,\ldots,x_n)
=
x_0  \odot x_1 \odot \ldots \odot x_n.
$$
\end{proposition}
\emph{Proof.} We will prove this by induction. For $n=1$   there is nothing to prove.
Assuming formula's validity for $n$ and defining $y_i=x_i \odot x_{i+1}$, we obtain
\begin{gather*}
f^{n+1}(x_0,x_1,\ldots,x_{n+1})=
f^{n}(y_0, y_1, \ldots, y_n)\\
=(x_0 \odot x _1) \odot (x_1 \odot x _2) \odot \ldots  \odot (x_{n} \odot x _{n+1})=\\
x_0 \odot (x _1 \odot x_1 )\odot (x _2 \odot x_2) \odot \ldots  \odot 
(x_{n} \odot x _{n}) \odot x_{n+1} \\
=x_0  \odot x_1 \odot \ldots \odot x_{n+1},
\end{gather*}
where we used idempotency and associativity. This proves validity of the formula
for $n+1$, as required. $\square$

For commutative semigroups one can obtain a similar result. 
We will use the following notation,
$$
x^{\cdot n} = \underbrace{x \odot x \odot \ldots \odot x}_n.
$$
This is to distinguish  $x^{\cdot n}$ from regular arithmetic powers $x^n$, as both will be used in the next section of this  paper. 
\begin{proposition}\label{mainprop}
If the semigroup $({\cal A},\odot)$ is commutative and $f(x_0,x_1)=x_0 \odot x_1$, then 
$$f^n(x_0,x_1,\ldots,x_n)
=
x_0^{\cdot \binom{n}{0}} \odot x_1^{\cdot \binom{n}{1}} \odot \ldots \odot x_n^{\cdot \binom{n}{n}}.
$$
\end{proposition}
\emph{Proof.} We will again use induction. The $n=1$ case is obvious.
Assume that the formula is true for $n$ and consider
$$f^{n+1}(x_0,x_1,\ldots,x_{n+1})=
f^{n}(y_0, y_1, \ldots, y_n),$$
where
$y_i=x_i \odot x_{i+1}$. Using the formula for $n$ we have
\begin{gather*}
f^{n}(y_0, y_1, \ldots, y_n)
=(x_0\odot x_1)^{\cdot \binom{n}{0}} \odot (x_1\odot x_2) ^{\cdot \binom{n}{1}} 
\odot \ldots \odot (x_{n-1} \odot x_n)^{\cdot \binom{n}{n}}\\
=x_0^{\cdot \binom{n}{0}} \odot x_1^{\cdot \binom{n}{0} + \binom{n}{0}} \odot 
x_2^{\cdot \binom{n}{0} + \binom{n}{2}}
\ldots 
\odot x_{n-1}^{\cdot \binom{n}{n-1} + \binom{n}{n}}  
\odot x_n^{\cdot \binom{n}{n}}.
\end{gather*}
Note that we used $(x \odot y)^{\cdot m}=x^{\cdot m} \odot y^{\cdot m}$ which is valid
assuming commutativity. Now, since
$$
\binom{n}{i} + \binom{n}{i+1}=\binom{n+1}{i+1}  \quad \text{and}\quad \binom{n}{0}=\binom{n+1}{0}=\binom{n}{n}=\binom{n+1}{n+1}=1,
$$
we obtain
\begin{gather*}
f^{n+1}(x_0,x_1,\ldots,x_{n+1})
=x_0^{\cdot \binom{n+1}{0}} \odot 
x_1^{\cdot \binom{n+1}{1}} \odot 
x_2^{\cdot \binom{n+1}{2} }
\ldots 
\odot x_{n-1}^{\cdot \binom{n+1}{n}}  
\odot x_n^{\cdot \binom{n+1}{n+1}},
\end{gather*}
proving validity of our formula for $n+1$, as required. $\square$

Note that the commutativity is not assumed in  Proposition \ref{thmidem},
thus Proposition \ref{thmidem} is \emph{not}  a special case of Proposition \ref{mainprop}.

\section{Solutions}
Having the results of the previous section, we can now construct solution formulae for all
18 two-input ternary rules. We will start with the simplest ones.
\subsection{Trivial rules}
Let us consider two trivial semigroups first, namely $G_7$ and $G_{17}$.
For $G_7$ we have $f_j(u,v)=2$, hence 
$f_7^n(x_0,x_1, \ldots, x_n)=2$. For $G_{17}$, $f_{17}(u,v)=u$, so $f_{17}^n(x_0,x_1, \ldots, x_n)=x_0$.  Therefore,
\begin{equation*}
[F^n_{7}(x)]_i =  2  \quad \text{and} \quad [F^n_{17}(x)]_i =  x_{i}.
\end{equation*}

Another rather simple case is $G_3$ for which  $f_{3} (u,v)=   2 u (u-2) v (v-2)$. This is a commutative rule, thus we can use Proposition~\ref{mainprop}. It is easy to verify that $x^{\cdot m}=0$ for  $x\in\{0,2\}$ when $m>1$. When $x=1$, we have
$1 \odot 1=2$. Since $2 \odot y=0$ for any $y$, this implies that $x^{\cdot m}=0$ for $x=1$ and $m>2$.
This yields

\begin{gather*}
f_3^n(x_0,x_1,\ldots,x_n)
=
x_0^{\cdot \binom{n}{0}} \odot x_1^{\cdot \binom{n}{1}} \odot \ldots \odot x_n^{\cdot \binom{n}{n}}
=\begin{cases}
x_0 \odot 0 \odot x_n & n>2, \\
x_0 \odot x_1^{\cdot 2} \odot x_2 &  n=2,\\
x_0 \odot x_1 &  n=1.
\end{cases}
\end{gather*}
Obviously $x_0 \odot 0 \odot x_n=0$. Furthermore, 
$x_0 \odot x_1^{\cdot 2} \odot x_2=0$ because $ x_1^{\cdot 2}$
is either 0 or 2 and $f(0,y)=0$, $f(2,y)=0$ for any $y$.
We also have
$x_0 \odot x_1= 2 x_0 (x_0-2) x_1 (x_1-2)$, therefore
$$
f_3^n(x_0,x_1,\ldots,x_n)
=\begin{cases}
2 x_0 (x_0-2) x_1 (x_1-2) & n=1, \\
0 &  n>1,
\end{cases}
$$
and finally
\begin{equation*}
 [F^n_{3}(x)]_i = \begin{cases} 
          2 x_{i} (x_{i} -2 ) x_{i+1} (x_{i+1}-2)    & n=1, \\
          0 & n>1 .
       \end{cases}
\end{equation*}

For $G_{11}$, the local function depends only on the first variable, 
$f_{11} (u,v)= 1+(u-1)^2$. We will define $g(u)=1+(u-1)^2$. Note that
$g(u)=g(g(u))$ for $u \in \{0,1,2\}$.
Since the semigroup is not commutative and it is not idempotent,  we cannot use any of our propositions. We can, however, iterate functions $f^n$ directly,
\begin{gather*}
{f}_{11}^n(x_0,x_1,\ldots,x_n)={f}_{11}^{n-1}(g(x_0), g(x_1), \ldots, g(x_{n-1}))\\
={f}_{11}^{n-2}(
g(g(x_0)), g(g(x_1)), \ldots, g(g(x_{n-2})))={f}_{11}^{n-2}( g(x_0), g(x_1), \ldots, g(x_{n-2}))\\
=\ldots ={f}_{11}(g(x_0), g(x_1))= g(x_0) = 1+(x_0-1)^2,
\end{gather*}
and finally
\begin{equation*}
[F^n_{11}(x)]_i =   1+(x_{i}-1)^2.
\end{equation*}
\subsection{Rules with solutions involving binomial coefficient}

A more interesting case are semigroups inducing CA rules   solvable by
direct application of Proposition \ref{mainprop}. The first one is
 $G_1$, for which $u\odot v=(u+v) \mod 3$, so Proposition \ref{mainprop} immediately yields
\begin{equation*}
[F^n_{1}(x)]_i =  \sum_{j=0}^{n} {\binom{n}{j}} x_{i+j}  \mod 3.
\end{equation*}
Another one is $G_4$, with multiplication operation given by  $u\odot v=1+(u-1)(v-1)$. It will be convenient to change variables to 
$\tilde{u}=u-1$ and $\tilde{v}=v-1$. This transforms the set of states to
$\{-1,0,1\}$ with semigroup multiplication defined as $\tilde{u} \odot
\tilde{v}=\tilde{u} \tilde{v}$. The semigroup multiplication in new variables is thus a regular arithmetic multiplication, and we obtain
$$
\tilde{x}_0^{\cdot \binom{n}{0}} \odot \tilde{x}_1^{\cdot \binom{n}{1}} \odot \ldots \odot \tilde{x}_n^{\cdot \binom{n}{n}}
=\prod_{j=0}^n \tilde{x_j}^{\binom{n}{j}}.
$$
Coming back to the original variables the solution formula follows immediately,
\begin{equation*}
[F^n_{4}(x)]_i =  1+\prod_{j=0}^{n} (x_{i+j}-1)^{\binom{n}{j}} .
\end{equation*}

The same method can be used for  $G_2$, with slight modification.
For this case we have $f_{2} (u,v)=  1+(v^2-v-1) (u^2-u-1)$ and
recall that $f_{4} (u,v)=   1+(v-1) (u-1) $.
It is easy to see  that $f_2(u,v)=f_4(g(u), g(v))$ where $g(x)=x^2-x$. 
Denoting by $g(x)$ a bi-infinite configuration such that $[g(x)]_i=g(x_i)$, this implies
$$
F_2(x)=F_4(g(x)).
$$
Now let us note that $g(x)=x^2-x$ takes values in
$\{0, 2 \}$ for any $x\in \{0,1,2\}$, hence
 $F_4(g(x))$ only contains 0's or 2's. Furthermore, one can easily verify that
 $$f_4(u,v)=f_2(u,v) \quad \text{for $u,v \in \{0, 2\}$}.
 $$
 For all further iterations, therefore, it does not matter if we apply $F_2$ or $F_4$, 
 meaning that
\textbf{\begin{equation*}
[F^n_{2}(x)]_i = [F_2^{n-1}F_4(g(x))]_i=  [F^n_{4}(g(x))]_i =
 1+\prod_{j=0}^{n} \big(g(x_{i+j})-1\big)^{\binom{n}{j}},
\end{equation*}}
and finally
\begin{equation*}
[F^n_{2}(x)]_i =  1+ \prod_{j=0}^{n}(x_{i+j}^2-x_{i+j}-1)^{\binom{n}{j}} .
\end{equation*}
For $G_6$ the derivation is the same, except that the  function $g$ is given by $g(x)=x^2-3x+2$, and we have
\begin{gather*}
f_6(u,v)= 1+(v^2-3 v+1) (u^2-3 u+1)\\=1+ \big(g(u)-1\big)\big(g(v)-1\big)
=f_4(g(u), g(u)).
\end{gather*}
Identical procedure as for $G_2$ above  yields the final result
\begin{equation*}
[F^n_{6}(x)]_i =      1+\prod_{j=0}^{n} (x_{i+j}^2-3 x_{i+j}+1)^{\binom{n}{j}}. 
\end{equation*}

Closely related to $G_6$ is another commutative semigroup $G_5$. 
Using the formulas for $f_5$ and $f_6$ from Table \ref{tablelocfunct} one can easily demonstrate that  
 $$f_{5} (u,v)=  f_6(u,v)-u(2-u)v(2-v),$$
 meaning that $f_5$ and $f_6$ differ only  when $u=v=1$.
Furthermore, $f_5(1,u)=f_5(u,1)=u$, which means that the pair 11 cannot appear in iterations of
rule $f_5$ if it was not present in the initial configuration. Therefore, if
the sequence $x_0,x_1, \ldots, n_1$ does not contain any 1's, 
we have
$$f_5^n(x_0,x_1,\ldots,x_n)=f_6^n(x_0,x_1,\ldots,x_n).$$
What happens if  $x_0,x_1, \ldots, x_n$ contains some 1's (but not all)?
Due to commutativity we can freely permute $x_i$'s and thus assume that $x_0=1$
and $x_1 \neq 1$. Note that $f_5(1,v)=f_6(2,v)$ if $v\neq 1$, and therefore 
\begin{gather*}
f_5^n(1,x_1,\ldots,x_n)=f_6^n(2,x_1,\ldots,x_n).
\end{gather*}
If there is more than a single 1, we can repeat this process again, bringing
this 1 to the first position and replacing it by 2. This can be repeated
for all 1's -- as long as there exist at least one entry different from 1.
Note that in the absence of 1's,  both $f_5$ and $f_6$ agree, meaning that
$$f_5^n(x_0,x_1,\ldots,x_n)=f_6^n(x_0,x_1,\ldots,x_n),$$
as long as there is at least one $i$ for which $x_i \neq 1$. When 
$x_i=1$ for all $i$'s, we have
$f_5(1,1,\ldots, 1)=1$ and $f_6(1,1,\ldots, 1)=2$, thus we can write
$$
f_5^n(1,x_1,\ldots,x_n)=
\begin{cases}
f_6^n(2,x_1,\ldots,x_n) -1 &  \text{if $x_1=x_2=\ldots =x_n=1,$} \\
f_6^n(2,x_1,\ldots,x_n) & \text{otherwise.} 
\end{cases}
$$
All we need to do now is to construct indicator function for $11\ldots 1$, which is
$$
1 - \prod_{j=0}^{n} x_{j}  (2-x_{j})  
=\begin{cases}
1 &  \text{if $x_1=x_2=\ldots =x_n=1$,} \\
0 & \text{otherwise.} 
\end{cases}
$$
Subtracting this indicator function from the expression for $F_6^n$ yields
\begin{equation*}
[F^n_{5}(x)]_i =   1+\prod_{j=0}^{n} (x_{i+j}^2-3 x_{i+j}+1)^{\binom{n}{j}}   - \prod_{j=0}^{n} x_{i+j}  (2-x_{i+j}).
\end{equation*}

\subsection{Rules 8 and 9}
For $G_8$, $f_{8} (u,v)=   u (2-u) v (2-v) = g(u)g(v)$, where
$g(u)=u(2-u)$.  When $u,v \in \{0,1\}$, $f_{8} (u,v)=uv$. Furthermore, 
$g(u)\in \{0,1\}$ for any $u\in \{0,1,2\}$ and $g(0)=0$, $g(1)=1$.
This means that 
$$x \odot x = (g(x))^2=g(x),$$
$$x \odot x \odot x = x \odot g(x) = g(x) g(g(x))=g(g(x))^2=g(x),$$
and by induction
$$x^ {\cdot m} = g(x).$$
This yields
\begin{gather*}
f^n(x_0,x_1,\ldots,x_n)
=
x_0^{\cdot \binom{n}{0}} \odot x_1^{\cdot \binom{n}{1}} \odot \ldots \odot x_n^{\cdot \binom{n}{n}}
=g(x_0) \odot g(x_1) \odot \ldots \odot g(x_n)\\
=\prod_{j=0}^n g(x_j)=\prod_{j=0}^n x_j(2-x_j),
\end{gather*}
and finally
\begin{equation*}
[F^n_{8}(x)]_i =  \prod_{j=0}^{n} x_{i+j} (2-x_{i+j}).  
\end{equation*}

For $G_9$, $f_{9} (u,v)=   \frac{1}{4} u v (3-v) (3-u)=g(u)g(v)$, where
$g(u)=\frac{1}{2}u(3-u)$. The function $g$ has the same properties as in the case of $G_8$, thus we will not repeat the derivation, writing just the final answer,
\begin{equation*}
[F^n_{9}(x)]_i =   \prod_{j=0}^{n} \frac{1}{2} x_{i+j} (3-x_{i+j}).
\end{equation*}

\subsection{Rule 10}
Rule 10 has the local function
$f_{10} (u,v)= u v (u+v-u v)$. We first note that 
$$
f_{10}(u,u)=u^3(2-u) = \begin{cases}
0 & u=0,\\
1 & u=1,\\
0 & u=2.
\end{cases}
$$
This implies that, for $n>1$, $x^{\cdot n}=0$ if $x \in \{0,2\}$ and $x^{\cdot n}=1$ if $x=1$.
Using the indicator function for 1 we can write
$$
x^{\cdot n}= \begin{cases}
x(2-x)& \text{if $n>1$},\\
x & \text{if $n=1$}.
\end{cases}
$$
Since $G_{10}$ is commutative, we have
\begin{gather*}
f_{10}^n(x_0,x_1,\ldots,x_n)
=
x_0^{\cdot \binom{n}{0}} \odot x_1^{\cdot \binom{n}{1}} \odot \ldots \odot x_n^{\cdot \binom{n}{n}}.\\
=x_0 \odot x_1(2-x_1) \odot x_2(2-x_2) \odot \ldots \odot x_{n-1}(2-x_{n-1}) \odot x_n.
\end{gather*}
Note that $x(2-x)$ takes only values 0 or 1, and if $u, v \in \{0, 1\}$ then $f_{10}(u,v)=uv$, i.e., the semigroup multiplication becomes just the normal arithmetic multiplication. For this reason,
\begin{gather*}
f_{10}^n(x_0,x_1,\ldots,x_n)
=x_0 \odot \left( \prod_{j=1}^{n-1}  x_j(2-x_j) \right) \odot x_n\\
=  \left( \prod_{j=1}^{n-1}  x_j(2-x_j) \right) \odot (x_0 \odot x_n).
\end{gather*}
The only possible values of the big product are 0 or 1, and only when it is equal to 1 
there is a possibility that the entire expression becomes non-zero. Furthermore,
$1 \odot x_0 \odot x_n= x_0 \odot x_n= x_0 x_n (x_0+x_n-x_0 x_n) $, yielding
$$
[F^n_{10}(x)]_i = x_{i} x_{i+n} (x_{i}+ x_{i+n}-x_i x_{i+n}) \prod_{j=1}^{n-1}  x_{i+j}(2-x_{i+j}).
$$

\subsection{Rule 12}
Rule 12 with the local function $f_{12} (u,v)= u v (2-u)$
has the property $f(0,v)=f(v,0)=0$. This means that if 0 is present in 
$x_0,x_1,\ldots, x_n$ then  $$f_{12}^n(x_0,x_1,\ldots, x_n)=0.$$ To get non-zero
output of $f_{12}$, therefore, its arguments must include only 1's and 2's. However,
$f(2,1)=f(2,2)=0$, meaning that the presence of pairs 21 or 22 in $(x_0,x_1,\ldots, x_n)$
will also cause the output to be zero. This leaves
 $1,1,\ldots, 1$ and $1,1,\ldots, 1,2$ as the only configurations producing
non-zero output, and it is easy to check that
$$
f_{12}^n(1,1,\ldots, 1)=1 \quad \text{and} \quad f_{12}^n(1,1,\ldots, 1,2)=2.
$$
 Indicator function of $(x_0,x_1,\ldots, x_{n-1})=(1,1,\ldots,1)$
is $\prod_{j=0}^{n-1}x_j (2-x_j)$, and if we multiply it by $x_n$, we will get the
desired result,
$$f_{12}^n(x_0,x_1,\ldots, x_n)=x_n\prod_{j=0}^{n-1}x_j (2-x_j),$$
hence
\begin{equation*}
[F^n_{12}(x)]_i =    x_{i+n} \prod_{j=0}^{n-1} x_{i+j} (2-x_{i+j}).  
\end{equation*}

\subsection{Idempotent rules 13--16}

Rule 13 has the local function $f_{13}=\max{(u,v)}$. It is both commutative and idempotent,
thus we can write
$$f_{13}^n(x_0,x_1,\ldots,x_n)
= x_0  \odot x_1 \odot \ldots \odot x_n.$$
Due to commutativity, we can sort $x_i$'s on the right hand side  of the above in an increasing order and then use
idempotency. If all three symbols $0,1,2$ are present among them, we will get
$$f_{13}^n(x_0,x_1,\ldots,x_n)=0 \odot 1 \odot 2 =2.$$ If only two different symbols $a$ and $b$ are 
present among $(x_0,x_1, \ldots, x_n)$, we get $$f_{13}^n(x_0,x_1,\ldots,x_n)=a \odot b =\max{(a,b)}.$$
In all cases, therefore, $f_{13}^n$ will return the largest value of its arguments,
$$f_{13}^n(x_0,x_1,\ldots,x_n)= \max {(x_0,x_1,\ldots,x_n)},$$
and
\begin{equation*}
[F^n_{13}(x)]_i = \max \{x_{i+j}, j=0,1,\ldots,n\} .  
\end{equation*}
Note that one could also use polynomial representation of $f_{13}$. Although is it rather
unwieldy, namely 
$f_{13}(u,v)=u+v-\frac{5}{2} u v-\frac{1}{2} u^2 v^2+u^2 v+u v^2$,
polynomial expression for $[F^n_{13}(x)]_i$ could be obtained as well. We will not pursue this here because the expression obtained above is much simpler.

For rule 14, $f_{14} (u,v)=  \frac{1}{2} u v (3 u v-5 u-5 v+9)$.
Since it is idempotent,
$$f_{14}^n(x_0,x_1,\ldots,x_n)
=
x_0  \odot x_1 \odot \ldots \odot x_n.
$$
This rule is also commutative. Because of this and because $0 \odot v=0$ for any $v$, 
if 0 appears in $x_0,x_1, \ldots, x_n$, then $f_{14}^n(x_0,x_1,\ldots,x_n)=0$.
We also have $1 \odot 2=0$, which means that  the only possibility
to obtain non-zero value of $f_{14}^n(x_0,x_1,\ldots,x_n)$ is to have
$x_0,x_1, \ldots, x_n$ consisting of all 1's or all 2's.
The indicator functions for all 1's and all 2's are, respectively,
$\prod_{j=0}^{n} x_{j} (2-x_{j})$ 
and
 $\prod_{j=0}^{n} \frac{1}{2}x_{i+j} (x_{i+j}-1)$, which immediately yields the solution formula 
\begin{equation*}
[F^n_{14}(x)]_i = \prod_{j=0}^{n} x_{i+j} (2-x_{i+j})  +2 \prod_{j=0}^{n} \frac{1}{2}x_{i+j} (x_{i+j}-1).         
\end{equation*}

Rule 15 is idempotent and non-commutative, with $f_{15} (u,v)=   \frac{1}{2}u (u v^2-u v-v^2-2 u+v+4)$. We have $f_{15}(0,v)=0$ or $0 \odot v=0$, hence
$$0  \odot x_1 \odot \ldots \odot x_n =
0  \odot x_2 \odot \ldots \odot x_n = \ldots = 0 \odot x_n =0.$$
Moreover, since  also $1 \odot v=1$, then by the same token
$$1  \odot x_1 \odot \ldots \odot x_n =1.$$
Let us now observe that $f_{15}(2,v)=v(v-1) \in \{0,2\}.$ This means that
if $x_0,x_1, \ldots, x_n$ starts with 2, the only possibility for
$f^n_{15}( x_0,x_1, \ldots, x_n)$ to return non-zero value is that it
consists of all 2's. We thus have
$$f^n_{15}( x_0,x_1, \ldots, x_n)=
\begin{cases}
1 & \text{if $x_0 =1$},\\
2 & \text{if $(x_0,x_1, \ldots, x_n)=(2,2, \ldots, 2)$},\\
0 & \text{otherwise}.
\end{cases}
$$  
The indicator function for $x_0=1$ is $x_0(2-x_0)$, and for all 2's it is
$\prod_{j=0}^n \frac{1}{2}x_j(x_j-1)$. This yields
$$
f^n_{15}( x_0,x_1, \ldots, x_n)=x_0(2-x_0)+ 2\prod_{j=0}^n \frac{1}{2}x_j(x_j-1),
$$
and hence
\begin{equation*}
[F^n_{15}(x)]_i =     x_{i} (2-x_{i})+ 2 \prod_{j=0}^{n} \frac{1}{2}x_{i+j} (x_{i+j}-1).  
\end{equation*}

Another idempotent and non-commutative semigroup is $G_{16}$ with the local function
$f_{16} (u,v)= 1+(u-1)(v-1)^2 $. Straightforward  calculations   
verify that   
\begin{align*}
x_0 \odot x_1                    &=1+(x_0-1)(x_1-1)^2,  \\
x_0 \odot x_1 \odot x_2          &=1+(x_0-1)(x_1-1)^2 (x_2-2)^2,  \\
x_0 \odot x_1 \odot x_2 \cdot x_3&=1+(x_0-1)(x_1-1)^2 (x_2-1)^2(x_3-1)^2, 
\end{align*}
etc. One can thus show by induction that
$$
x_0  \odot x_1 \odot \ldots \odot x_n
= 1+ (x_0-1)\prod_{j=1}^n (x_j-1)^2,
$$
hence
\begin{equation*}
[F^n_{16}(x)]_i =  1+ (x_{i}-1) \prod_{j=1}^{n}  (x_{i+j}-1)^2.
\end{equation*}
\subsection{Rule 18}
Rule 18 with $f_{18} (u,v)=  \frac{1}{2} u (u v-2 u-v+4)$ is non-commutative
but idempotent, so Proposition \ref{thmidem} can be used.
First note that $0 \odot v =0$, $1 \odot v =1$, and $2 \odot v =v$.  
Using this and associativity,
we can easily see that
\begin{align*}
&0 \odot \ldots =0,
&&1 \odot \ldots = 1,\\
&2 \odot 0  \odot  \ldots=0,
&&2 \odot 1 \odot \ldots = 1,\\ 
&2 \odot 2 \odot 0  \odot  \ldots=0,
&&2 \odot 2 \odot 1 \odot \ldots = 1,\\ 
&2 \odot 2 \odot 2 \odot 0  \odot  \ldots=0,
&&2 \odot 2 \odot 2 \odot 1 \odot \ldots = 1,\\
&\ldots &&\ldots  \\
&\underbrace{2 \odot 2 \odot \ldots \odot   2}_{n} \odot 0  =0,
&&\underbrace{2 \odot 2 \odot \ldots \odot   2}_{n} \odot 1  =1.
\end{align*}

It is clear that the above list contains all possible configurations of $(x_0, x_1, \ldots, x_n)$ except the configuration consisting of all 2's, for which we have
$$
\underbrace{2 \odot 2 \odot \ldots \odot   2}_{n+1}=2.
$$
This means that the only possibility to produce 1 as output of $f^n_{18}$ is to
have its arguments starting with 1 or with a block of 1's terminated by 2. The only possibility to get output 2 is to have all 2's as input. Define indicator function
\begin{equation*}
H(k) = 
\begin{cases}
x_0 (2-x_0) & k=0,\\
\displaystyle x_{n} (2-x_{n})   \prod_{j=0}^{n-1} \frac{1}{2}x_{j} (x_{j}-1) & k>0.
\end{cases}
\end{equation*}
We can see that $H(k)=1$ if the configuration starts with 1 and $k=0$.
If, for $k>1$,  it starts with $k$ 1's terminated by 2, we have $H(k)=1$ as well. In all other cases $H(k)=0$. In order to obtain 1 as output of $f^n_{18} (x_0, x_1, \ldots , x_n)$ we want
$$H(0)=1\text{\,\, or \,\,} H(1)=1  \text{\,\, or \,\,} \text{\,\, $\ldots$ \,\,} \text{\,\, or \,\,}
H(n)=1.
$$
Since in Boolean logic $p \vee q = \sim(\sim p\,\wedge \sim q)$, the above is equivalent to
$$
1- \prod_{k=0}^n \left(1- H(k) \right) =1.
$$
Indicator function of all 2's is
$$
\prod_{j=0}^{n} \frac{1}{2} x_{j} (x_{j}-1),
$$
and we need to multiply it by 2 to get the correct value of the output.
Combining all of this together we obtain
$$f^n_{18}(x_0,x_1, \ldots, x_n)
=1- \prod_{k=0}^n \left(1- H(k) \right) + 2 \prod_{j=0}^{n} \frac{1}{2} x_{j} (x_{j}-1).
$$
The final solution formula, therefore,  can be written as
\begin{equation*}
[F^n_{18}(x)]_i =1-\prod_{k=0}^{n} (1-H(i,k) ) + 2 \prod_{j=0}^{n} \frac{1}{2} x_{i+j} (x_{i+j}-1),
\end{equation*}
where
\begin{equation*}
H(i,n) = 
\begin{cases}
x_i (2-x_i) & n=0,\\
\displaystyle x_{i+n} (2-x_{i+n})   \prod_{j=0}^{n-1} \frac{1}{2}x_{i+j} (x_{i+j}-1) & n>0.
\end{cases}
\end{equation*}

\section{Concluding remarks}

We have demonstrated that all CA defined by semigroups with 3 elements are solvable. What are the possible uses of such solution formulae? 
Since they exhibit explicit dependence of the state of site $i$ at iteration $n$ on the initial
configuration, they can be used  to compute various quantities of interest along the orbit of the given initial configuration.

For example, let us assume that the initial configuration
is drawn from a Bernoulli distribution such that $Pr(x_i=k)=p_k$ for $k=0,1,2$, where
$p_0,p_1,p_2 \in [0,1]$ and $p_0+p_1+p_2=1$. By taking the expected value of both sides of the solution formula and by using basic properties of expected values, one can obtain \cite{paper57,book1}
the expected value of the state of site $i$ after $n$ iterations of the CA as a function of probabilities
$p_0,p_1,p_2$. Probabilities of occurrences of finite block of symbols can also be computed by a similar method \cite{book1}.
The solution formulae are thus very useful to study statistical properties of the corresponding CA.

Another possible use of the results presented here would be the investigation of finite size effects. If one takes
as the initial configuration a periodic configuration with period $L$, then it is equivalent to what is often called
``finite lattice with periodic boundary conditions''. As shown in \cite{book1}, one can use the solution formulae
to study how various properties of the orbit of such configuration depend on the size of the lattice $L$. The author
plans to report relevant results elsewhere in the near future.

\begin{credits}
\subsection{\ackname}
The author acknowledges comments made by anonymous referees.
Their thorough and careful reading of the manuscript helped to eliminate numerous typos and to correct the derivation of the formula for rule 3.
\end{credits}

\section*{Appendix: summary of solutions}
\small 
\begin{equation*}
[F^n_{1}(x)]_i =  \sum_{j=0}^{n} {\binom{n}{j}} x_{i+j}  \mod 3,
\quad
[F^n_{2}(x)]_i =  1+ \prod_{j=0}^{n}(x_{i+j}^2-x_{i+j}-1)^{\binom{n}{j}}, 
\end{equation*}
\begin{equation*}
 [F^n_{3}(x)]_i = \begin{cases} 
          2 x_{i} (x_{i} -2 ) x_{i+1} (x_{i+1}-2)    & n=1, \\
          0 & n>1, 
       \end{cases}
\quad
[F^n_{4}(x)]_i = 1+\prod_{j=0}^{n} (x_{i+j}-1)^{\binom{n}{j}}, 
\end{equation*}
\begin{equation*}
[F^n_{5}(x)]_i =   1+\prod_{j=0}^{n} (x_{i+j}^2-3 x_{i+j}+1)^{\binom{n}{j}}   - \prod_{j=0}^{n} x_{i+j}  (2-x_{i+j}),  
\end{equation*}
\begin{equation*}
[F^n_{6}(x)]_i =      1+\prod_{j=0}^{n} (x_{i+j}^2-3 x_{i+j}+1)^{\binom{n}{j}},  
\quad \quad
[F^n_{7}(x)]_i =  2,
\end{equation*}
\begin{equation*}
[F^n_{8}(x)]_i =  \prod_{j=0}^{n} x_{i+j} (2-x_{i+j}), 
\quad
[F^n_{9}(x)]_i =   \prod_{j=0}^{n} \frac{1}{2} x_{i+j} (3-x_{i+j}),  
\end{equation*}
\begin{equation*}
[F^n_{10}(x)]_i = x_{i} x_{i+n} (x_{i}+ x_{i+n}-x_i x_{i+n}) \prod_{j=1}^{n-1}  x_{i+j}(2-x_{i+j}),
\end{equation*}
\begin{equation*}
[F^n_{11}(x)]_i =   1+(x_{i}-1)^2,
\quad
[F^n_{12}(x)]_i =    x_{i+n} \prod_{j=0}^{n-1} x_{i+j} (2-x_{i+j}),  
\end{equation*}
\begin{equation*}
[F^n_{13}(x)]_i = \max \{x_{i+j}, j=0,1,\ldots,n\},   
\end{equation*}
\begin{equation*}
[F^n_{14}(x)]_i = \prod_{j=0}^{n} x_{i+j} (2-x_{i+j})  +2 \prod_{j=0}^{n} \frac{1}{2}x_{i+j} (x_{i+j}-1),         
\end{equation*}
\begin{equation*}
[F^n_{15}(x)]_i =     x_{i} (2-x_{i})+ 2 \prod_{j=0}^{n} \frac{1}{2}x_{i+j} (x_{i+j}-1),  
\end{equation*}
\begin{equation*}
[F^n_{16}(x)]_i =  1+ (x_{i}-1) \prod_{j=1}^{n}  (x_{i+j}-1)^2, 
\quad
[F^n_{17}(x)]_i =  x_{i},
\end{equation*}
\begin{equation*}
[F^n_{18}(x)]_i =1-\prod_{k=0}^{n} (1-H(i,k) ) + 2 \prod_{j=0}^{n} \frac{1}{2} x_{i+j} (x_{i+j}-1),
\end{equation*}
where
\begin{equation*}
H(i,n) = 
\begin{cases}
x_i (2-x_i) & n=0,\\
\displaystyle x_{i+n} (2-x_{i+n})   \prod_{j=0}^{n-1} \frac{1}{2}x_{i+j} (x_{i+j}-1) & n>0.
\end{cases}
\end{equation*}
\end{document}